\documentclass[10pt, a4paper, notitlepage]{article}

\usepackage{tgpagella}  
\usepackage{mathpazo}   

\usepackage{geometry}
\usepackage{amsmath}
\usepackage{graphicx}
\usepackage[numbers]{natbib}
\bibpunct{(}{)}{;}{a}{,}{,}
\usepackage[colorlinks=true, linkcolor=black,citecolor=black,urlcolor=black]{hyperref}
\usepackage{url}
\usepackage{breakurl}

\usepackage{sectsty}
\sectionfont{\fontsize{10}{10}\selectfont}
\subsectionfont{\fontsize{10}{10}\selectfont}

\usepackage{enumitem}

\usepackage{endnotes} 
\let\footnote=\endnote

\let\origfootnote\footnote%
\renewcommand{\footnote}[1] {%
\renewcommand\footnotesize\scriptsize%
\linespread{1.2}%
\origfootnote{#1}%
}%

\usepackage{fancyhdr}
\begin{document}

\title{\bfseries\Large Updating the Wassenaar Debate Once Again: \\ Surveillance, Intrusion Software, and Ambiguity \vspace{30pt}\\}

\author{Jukka Ruohonen \vspace{3pt}\\
\normalsize Department of Future Technologies \\
\normalsize University of Turku, Finland 
\vspace{20pt}\\
Kai K.~Kimppa \vspace{3pt}\\
\normalsize Department of Management and Entrepreneurship \\
\normalsize Turku School of Economics, Finland \\
}

\date{}
\maketitle
\thispagestyle{empty}

\begin{abstract}
This paper analyzes a recent debate on regulating cyber weapons through multilateral export controls. The background relates to the amending of the international Wassenaar Arrangement with offensive cyber security technologies known as intrusion software. Implicitly, such software is related to previously unregulated software vulnerabilities and exploits, which also make the ongoing debate particularly relevant. By placing the debate into a historical context, the paper reveals interesting historical parallels, elaborates the political background, and underlines many ambiguity problems related to rigorous definitions for cyber weapons. Many difficult problems remaining for framing offensive security tools with multilateral export controls are also pointed out.
\vspace{25pt}\\
\begin{center}
export controls, nonproliferation, Internet governance\vspace{35pt}\\
\end{center}
\begin{center}
\textit{\normalsize Journal of Information Technology \& Politics, \vspace{3pt}\\ published online in May 2019}
\vspace{25pt}\\
\small This is the authors' copy. The publisher's copy is available online at: \vspace{3pt}\\ \url{https://doi.org/10.1080/19331681.2019.1616646}
\end{center}
\end{abstract}

\thispagestyle{empty}
\clearpage
\setcounter{page}{1}

\section{Introduction}

In 2015 the United States (U.S.) Department of Commerce announced a proposal for an implementation of the amendments that were made in 2013 to the international Wassenaar Arrangement \citeyearpar{Wassenaar19a} on conventional weapons and related technologies that may be used for military purposes. The proposal addressed a new type of cyber weapons known as intrusion software, causing a vocal protest in the multinational information and communication technology (ICT) sector in general \citep{Shaw16}, and the cyber security industry in particular \citep{Burkart17}.\footnote{~For the purposes of this paper, the concept of cyber weapons is deliberately left undefined (for a theoretical discussion about the concept see \citealt{BellovinLandau17} and \citealt{EilstrupSangiovanni18}).} This paper sets to elaborate the debate from a historical perspective.

The rationale and relevance are easy to justify. From the beginning, Wassenaar was strongly motivated by and built upon the ideals of efficiency, openness, transparency, and responsibility \citep{Atlas08, Boese98, CraftGrillot99, Micara12, vonderDunk09}. In light of these ideals, the current debate exemplifies many of the pressing issues that continue to plague export controls. Adverse to the stated goals, the amendments and proposals were made without extensive prior public discussion and debate. A discussion is required also in academic research because academic research itself may be affected. In other words, the continuing alterations made to export controls may have important implications not only for the private sector cyber security industry but also for academic computer science research of computer security. As has long been the case with export controls for cryptography and science in general~\citep{Bohm99, Evans12, WilliamsJones14}, cross-border research and development, education, and the freedom of speech might all be exposed to the cyber security policy changes and initiatives elaborated.

This paper proceeds by discussing the historical background in terms of international venues and historical ideas, ideologies, and legacies, including the the 1990s debates on export controls for cryptography. By contrasting these against the current Wassenaar debate, the paper reveals a number of interesting historical parallels and differences. 

For framing the historical background, the paper emphasizes the concept of \textit{discretion}, which can be defined in the present context as a lack of obligatory rules for a state to implement and enforce a transfer control in a \textit{multilateral} export regulation regime covering physical goods, technology, or services. Here, the specific term multilateral is used interchangeably with a term \textit{inter-governmental}. Then: if a level of discretion is high -- as is the case with Wassenaar -- the efficiency of a given multilateral export control system depends on an agreement between the member states on the content, comprehensiveness, diligence, and transparency of export control lists agreed, which are, however, left for states to implement and enforce in order to prevent a transfer of a listed item to the hands of states targeted \citep{CraftGrillot99}.\footnote{~This kind of voluntarism has also remained Wassenaar's central problem \citep{Gartner10}. In the contemporary global world, an export to Middle East might be denied for a supplier in the U.S., although a transfer might be possible via a subsidiary in another country, to paraphrase \citeauthor{Beck00}'s \citeyearpar{Beck00} example. Exporters' failures to comply are relatively common \citep{Burke12}. Inadequate comprehensiveness and lack of diligence have frequently also opened different loopholes through which \textit{shadow exports} are still possible despite of placed controls~\citep{Waltz07}. The ICT sector is supposedly particularly problematic in this regard.} Due to the ideals of comprehensiveness and diligence, there is a continuing race for export control regimes to keep up with technical innovations and new forms of security, which should be enumerated meticulously for efficiency and bargaining reasons. 

At the same time, the discretion principle makes implementation a national or a regional matter, which implies that enforcement is exposed to national politics and policies, including economic policy. Consequently, many fundamental analytical pairs, such as freedom and security, benevolent and malevolent, and legitimate and illegitimate \citep{Rath14, WilliamsJones14}, are arguably not enough for understanding the historical background. In many respects, the export control history regarding ICT is better read against the pairs between security and commerce, foreign policy and trade policy, or more concretely, as an interplay between the U.S.~Departments of State and Commerce (cf.~\citealt[][]{DiffieLandau07, Evans12, Hansen16, Seyoum17}). This economic viewpoint also allows to better understand the current debate and its linkages to contemporary security industry practices and tools.

In addition to contributing to the scholarly research of export controls, this paper participates to the ongoing discussion about the increasing pressure for regulating cyber weapons, critical infrastructure, and related context areas of contemporary cyber security politics. In particular, the current Wassenaar debate is related to potential regulative elements placed upon \textit{software vulnerabilities} (software defects that expose software implementations to security weaknesses) and \textit{exploits} (software implementations that target vulnerabilities and with which target software implementations are potentially compromised). It can be noted that neither concept is explicitly related to \textit{malware}, which includes software implementations that are installed as payloads to systems that have already been compromised with exploits.\footnote{~This threefold terminological setup is not comprehensive, but it still allows framing the scope of the paper as well as the current debate. Importantly, the restriction to these three concepts deliberately bypasses arguably more fundamental concepts related to threats, risks, and likelihoods. Nor is the attempt to place the case at hand into analytical cages such as the confidentiality, integrity, and availability (CIA) triad.} The point is important from a regulative perspective. 

Engineering and distributing \text{malware -- whether} computer viruses or more recent nuisances such as \textit{ransomware} -- are already criminal offenses in many national legislations. In contrast, regulations for vulnerabilities and exploits are few and far between, although recent regulative pressures have also transcended to scholarly research and policy recommendations. Even though some scholars have been cautious to give policy recommendations \citep{Herr16,Wilson13}, many scholars have taken a firm stance recently in favor  \citep{Cavelty14, Jardine15} or against \citep{Denning15} the regulation of vulnerabilities and exploits. This paper contributes to this discussion with a case that alarms policy makers and stakeholders about many problems for regulation via export controls. Based on a brief technical review, a few policy recommendations are also pointed out.

To better understand the background, it should be emphasized that a software vulnerability is only an abstraction for a software bug with particular importance. Not all vulnerabilities are exploitable as such, and a concrete demonstration of the existence of a vulnerability is best done with a \textit{proof-of-concept} exploit. Although both concepts thus generalize to common software engineering abstractions, pressure to regulate these abstractions has had a more concrete manifestation, relating to a software security industry segment that emerged in the early 2000s for transacting sensitive information that vulnerabilities carry. 

These transactions are related to the practices, norms, and ideals behind the so-called \textit{vulnerability disclosure} via which benign vulnerability discoverers make their discoveries known to software vendors and the public~\citep{Buchanan16b}. Although different public sector organizations have participated in vulnerability disclosure for decades via national computer emergency response teams and related arrangements, the recent interest of policy makers and state agencies have focused on vulnerability and exploit marketplaces orchestrated by third-party brokerage companies, commonplace vulnerability finding contests and campaigns directly orchestrated by many ICT companies, and related commercial aspects that characterize particularly the offensive security industry segment \citep{Burkart17, Ruohonen16RCIS, Wilson13}.\footnote{~Here, \textit{offensive security} is used to analytically distinguish practices and technologies that are distinct to those used in the more traditional \textit{defensive security} segment, including conventional anti-virus software companies \citep{Ruohonen16RCIS}. This definition is deliberately restricted to the industry (for a more encompassing discussion see \citealt{EilstrupSangiovanni18} and \citealt{Slayton16}). Analogously to the duality between vulnerabilities and exploits -- often, demonstrating the existence of the former mandates the engineering of the latter, the (defensive) technologies within the sector often utilize offensive techniques for defensive purposes. The so-called \textit{penetration testing} \citep{Knowles16} is the prime example in this regard. Furthermore, it can be noted that the term \textit{zero-day} vulnerability refers to a vulnerability that has been discovered, whether by a benign actor or a criminal, but not made known (that is, disclosed) to the affected vendor, its users, and the public.} As shall be elaborated, however, the 2013 amendment to the Wassenaar Arrangement (WA) and the later implementation proposal of the U.S.~Department of Commerce both contained vagueness and ambiguity that are typical to export controls, which typically and often inconsequentially escalate to other domains.

\section{Background}

The current Wassenaar debate cannot be understood without understanding the historical background of modern expert controls. To present a concise historical narrative, the following discussion briefly addresses the international background and the historical genesis behind the Wassenaar Arrangement. These two facets are used to motivate the subsequently discussed early history of cyber security export controls.

\subsection{Export Control Regimes}

The Wassenaar Arrangement was initiated and signed between 1994 and 1995 as a replacement for the agreements Western states had had against the former Warsaw Pact member states. Driven by the stated goals of preventing the proliferation of weapons of mass destruction (WMD) and improving transfer regulation of conventional weapons, the WA was negotiated from the foundation that the the Cold War era Coordination Committee for Multilateral Export Controls (COCOM) had left after dissolving in 1994. Akin to a famous Dutch labor market agreement in 1982, the town of Wassenaar in the Netherlands lent its name for the final resolution known as the Wassenaar Arrangement of Export Controls for Conventional Arms and Dual-Use Goods and Technologies.

The loose and debatable term \textit{dual-use} refers to goods and technologies that are primarily commercial but may be still used for significantly enhancing military capabilities. In general, the term operates with the polar coordinates between civilian and military \citep{CraftGrillot99, Rath14}, although national legislations and norms add further coordinates. Often, national questions related to dual-use weapons are handled in specific national institutions, and these have usually also different export requirements and licensing procedures compared to conventional weapons and munitions \citep{DiffieLandau07, Seyoum17}.  This legal, bureaucratic, and political complexity is understandable against the often large amount of specific items covered in national and international regulations. Originally the list of multilaterally covered items ranged from tanks, large-caliber artillery,  and warships to electronics, semiconductors, lasers, and other manufactured military hardware, but international coverage was  later extended toward biological and chemical weapons, among other domains \citep{Atlas08, Shaw16}. Within this international and intergovernmental regulation and coordination setup, the WA has been the most important facilitating medium for items that fall within an all-embracing category of dual-use ICT.

\begin{figure*}[p!]
\centering
\includegraphics[width=\linewidth, height=8.0cm]{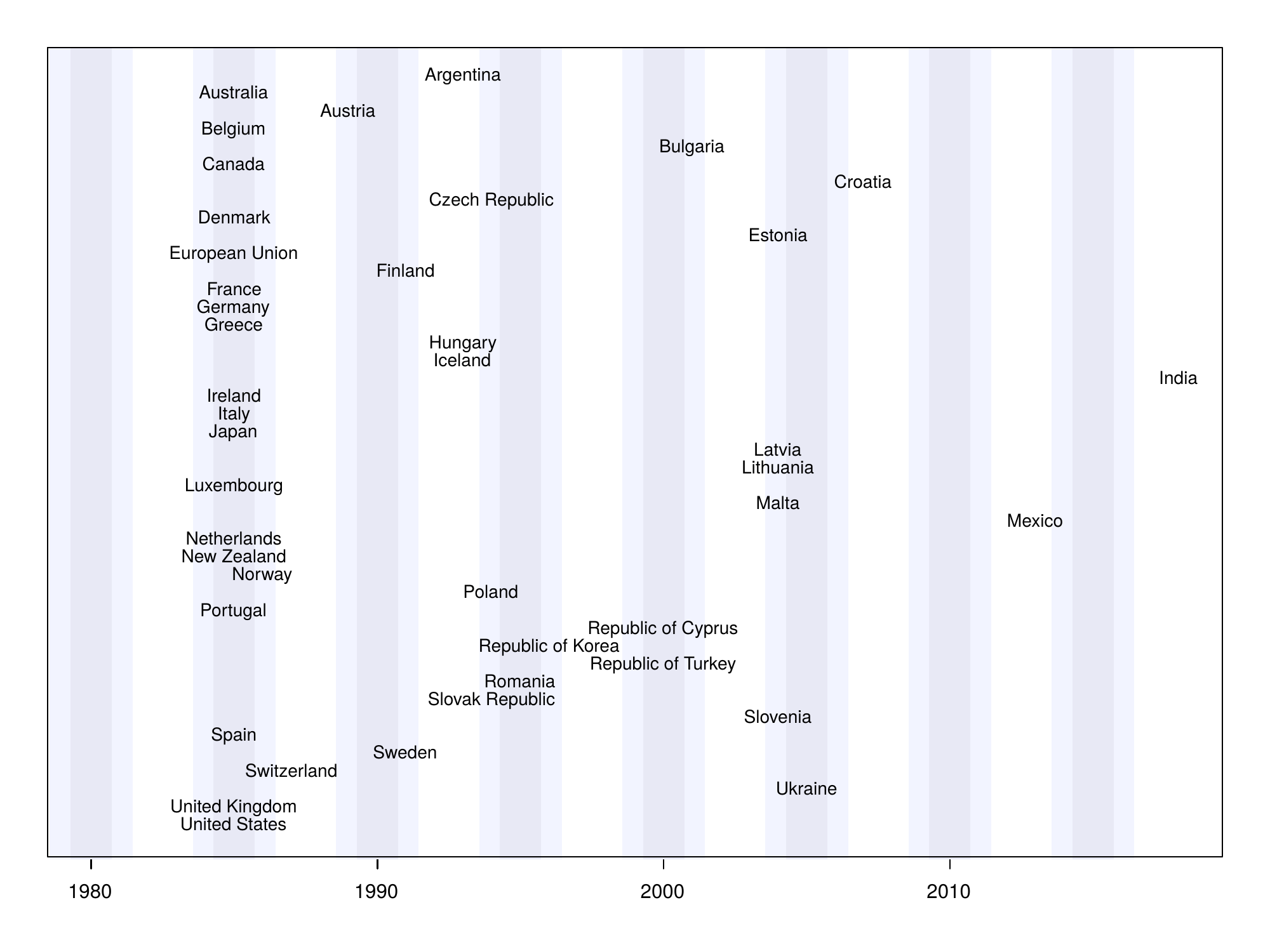}
\caption{\small The Australia Group Members (years of joining, as of 2019)}
\label{fig: australia}
%
\centering
\includegraphics[width=\linewidth, height=8.0cm]{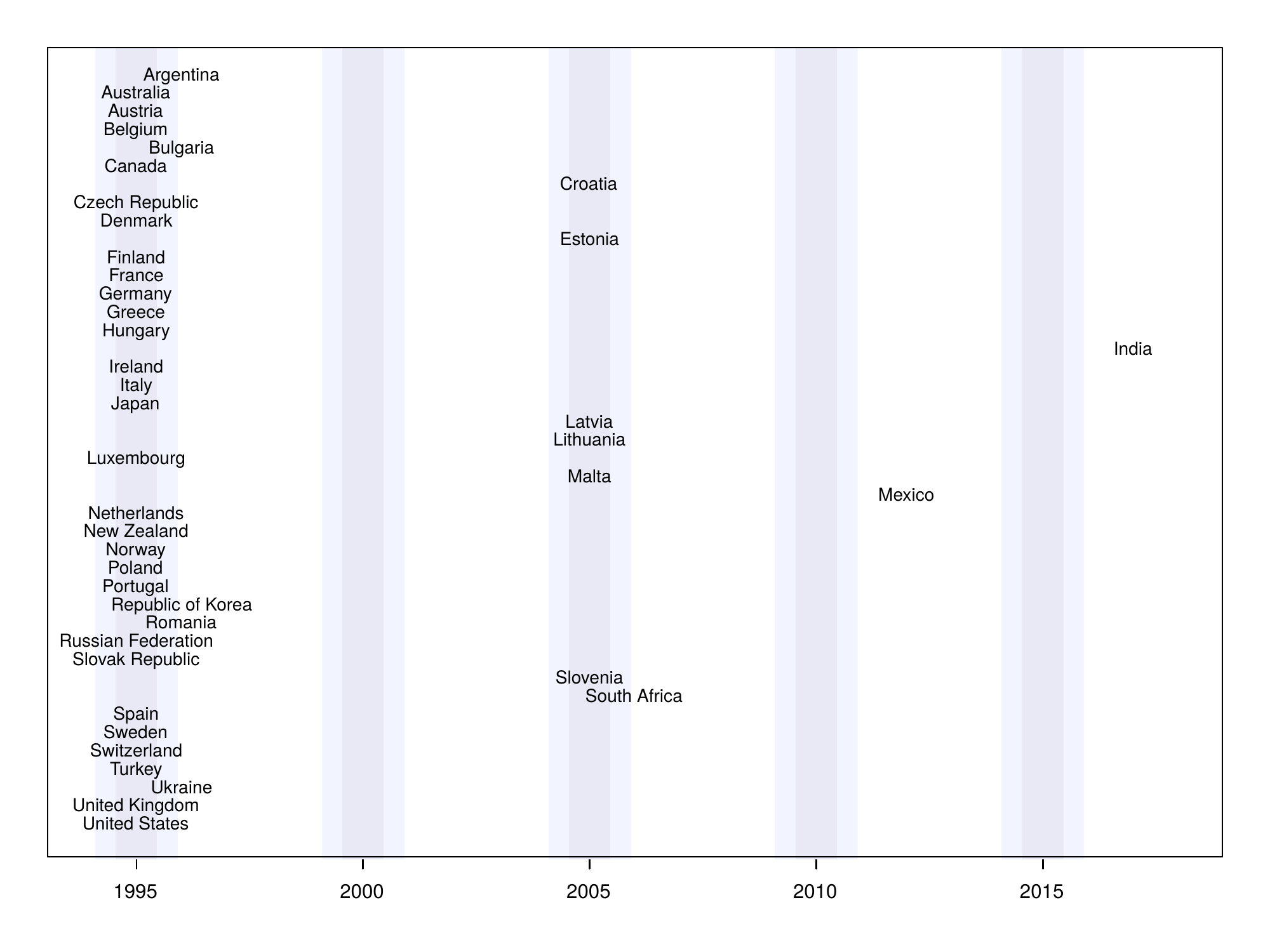}
\caption{\small The Wassenaar Arrangement Members (years of joining, as of 2019)}
\label{fig: wassenaar}
\end{figure*}

The expansion of coverage was fostered by international developments. A~brief calendar of noteworthy Cold War events includes the nuclear test initiated by India in 1974, followed by the later rocket tests by India and South Korea, the chemical warfare used during the Iran-Iraq war, and the late 1980s concerns of Iraq's state-led biological WMD programs \citep{Micara12, Shaw16}. These developments correlated with the formation of the Nuclear Suppliers' Group (NSG) in 1977, the Australia Group in 1985, and the Missile Technology Control Regime (MTCR) in 1987. From these, the Australia Group has continued to provide the premium but still informal venue for states to harmonize export controls of chemical and biological weapons. Currently, the composition of the Australia Group (see Fig.~\ref{fig: australia}) is almost identical with the composition of the Wassenaar Arrangement (see Fig.~\ref{fig: wassenaar}). Both multilateral venues have also continued to expand in the 2010s, both in terms of member states and covered items. Even though India joined in 2017, many important countries still remain unaffiliated, however. From the four key exceptions in the late 1990s~\citep{Boese98}, only South Africa has joined to the WA as an official member; Brazil, China, and Israel remain outside.

The partial expansion of the Wassenaar Arrangement toward conventional (non-WMD) weapons occurred already in the mid-1990s when the initial member states agreed to avoid transferring either weapons or ammunition to parties in the Afghanistan conflict, following the United Nations (UN) Security Council Resolution \citep{Wassenaar96, UnitedNations96}. After slower progression during the second half of the 1990s, fears of WMD proliferation again intensified in the aftermaths of the September 11 attacks  \citep{Micara12, WilliamsJones14}. In terms of Wassenaar, the changed security landscape after 2001 manifested itself by an alteration to the founding document in 2002, for the first time since the WA's initiation. Although pressing WMD proliferation questions were firmly on international political agenda also in the late 2000s and early 2010s, a major international breakthrough occurred with different kinds of weapons with the initiation of the  multilateral, UN-led, Arms Trade Treaty (ATT) program.

The focus on human rights was pronounced during the developments that led to the ATT program \citep{Bromley12, Erickson15}. This focus also motivated the 2013 amendments made to the WA. In particular, the initial impetus for the European Union (EU) and the U.S.~to restrict the proliferation of cyber weapons came with the numerous early 2010s cases that spotlighted the oppressive government use of surveillance technologies (\citealt{BauerBromley16, Bohnenberger17}; cf.~also \citealt{Amnesty14, HumanRightsWatch14}). As argued by \citet{McKuneDeibert17}, such technologies together with the briefly elaborated international background make it understandable why and how offensive cyber security technologies initially entered to the WA's realm: when policy makers approached cyber security by thinking about cyber weapons, the next logical step was to think also in terms of export controls for arms, weapons, and munitions. 

However, there are a couple of important points that further shed light on the export controls for cyber security technologies. The first point relates to a state-centric international relations viewpoint, which tends to conceal the long history of cyber security export controls and their relation to national politics. This point is particularly important for understanding the later 2015 proposal for implementing the WA in the United States. The second point stems from the discretion principle: the ideal of comprehensiveness provides a rationale for cataloging new security technologies, while voluntarism often makes it politically easy to add such technologies to the WA. Both points require a brief elaboration.

\subsection{Historical Genesis}

The international background can be used for illustrating a few key differences between Wassenaar and other comparable multilateral transfer control treaties and coordination bodies. When compared to the COCOM,  Wassenaar was from the beginning more heterogeneous, covering more states than the seventeen that had participated in the COCOM with its explicitly stated historical goal of preventing weapon exports to the Soviet block and China. The composition is an important characteristic also today. Given the annexation of Crimea by Russia in 2014, it is worthwhile to note that both Russia and Ukraine are participating in the Wassenaar Arrangement (see Fig.~\ref{fig: wassenaar}). If nothing else, this case exemplifies the commonly expressed criticism that the heterogeneous composition may hamper bargaining and consensus-building among the participating states \citep{Beck00, Bromley12}.\footnote{~Moreover, the current composition makes it possible to question whether terms such as Western or transatlantic are misleading for characterizing the case. While keeping this terminology remark in mind, in this paper, when used, these terms refer to the long historical genesis rather than to the geographic composition.} Yet, as \citet{CraftGrillot99} point out, already during the initiation, Wassenaar differed significantly from the COCOM not only in terms of the (a)~heterogeneous composition of the state members, but also with respect to (b) the lack of explicitly stated targets, (c)~the expansion of coverage to conventional weapons, (d)~and the omission of a mechanism for \textit{multilateral oversight}. The last two items are particularly relevant to the current cyber security debate.

The phrasing about lacking of multilateral oversight means that Wassenaar is built on national discretion; unlike with the historical COCOM, there are no formal reviews or veto points that the member states could use prior to a weapons transfer by another member state \citep{CraftGrillot99, Wassenaar96}. This Western-initiated historical genesis -- with its cartel-like constraints for the supply-side characteristics -- has also been met with skepticism in countries such as China, which has historically been reluctant to join to  the WA, MTCR, and Australia Group \citep{Yuan02}. Among the traditional Western allies of the U.S., on the other hand, the genesis has largely blended to national and regional legislations.

Although the EU has went slightly further in the coverage and enforcement among its member states, the regulations, including the core dual-use Council Regulation 428/2009 \citep{EC09a}, are strongly tied to the multilateral arrangements, including the NSG, MTCR, Australia Group, Wassenaar, and the UN treaties and resolutions \citep{Atlas08, Bohm99, Micara12, vonderDunk09}. To summarize: the COCOM and Wassenaar provided historical templates for constructing other multilateral venues~\citep[][]{Shaw16}, while the U.S.~export control tradition was often seen as a historical ``gold standard'' against which many of these venues were framed \citep{Erickson15, Waltz07}. The emphasis placed on discretion and commercial \textit{bona fide} transactions have been the central keywords in this transatlantic regime (\citealt{Bohm99}; see also \citealt{DiffieLandau07, vonderDunk09}). The generally liberal position \citep{Winkel03} on the controls for commercial weapons transfers together with the discretion principle make it understandable why this multi-treaty regime has often been used for both foreign and trade policies. Cyber security is no exception.

This genesis should be kept in mind before proceeding to the cyber security domain. In fact, within this domain in particular, the historical narrative reads pronouncedly as a story of the national U.S.~export control system. However, it should be also understood that, in reality, the U.S.~has not had a single unified system, policy, or strategy toward export controls after the Cold War. 

The occasional inconsistency and ambivalence have largely originated from national U.S.~politics \citep{Erickson15, Waltz07}. In the cyber security context, a basic political dividing line in the U.S.~has historically manifested itself between trade promoters and nonproliferation advocates, although, in general, the U.S.~cyber security politics and lobbying traits include numerous different institutions, companies, and non-governmental organizations (NGOs).\footnote{~By and large, this dividing line can be argued to apply also to Europe within which the public outcries for tighter controls have often met the commercial rationale of creating a level playing field for the European defense and security industries \citep{BauerBromley16, Hansen16}. In terms of the 2010s debate, however, the voice of the industry is largely a global one rather carrying a particular geographic tone.} Among the notable U.S.~export control institutions are the Department of State and the Department of Commerce, the latter having traditionally been the main institution responsible for questions related to dual-use weapons \citep{Burke12, DiffieLandau07, Rajeswari98, Seyoum17, vonderDunk09}. The former has traditionally concentrated on military technology, national security, and foreign policy together with institutions such as the Departments of Defense, Homeland Security, Justice, and illustratively for underlining the policy complexity, even the Departments of Treasury and Energy~\citep{Evans12, Rajeswari98, Rowold15, Seyoum17, Waltz07}.\footnote{~This kind of large institutional variety is not limited to bureaucratic administrative institutions (that is, the policy complexity extends to parliamentary institutions). Nor is it specific to the United States.~In general, poor coordination between institutions may also contribute to problems in rigorous enforcement, alongside other inefficiencies in export controls \citep{Seyoum17, Yuan02}.} Nevertheless, occasionally the fine balance between trade and security has been adjusted by concretely shifting export control responsibilities between the Departments of Commerce and State, respectively \citep{Thomsen01}. Against this backdrop, it was perhaps a small surprise that the current debate originated from a proposal made not via cyber security institutions but through the Department of Commerce. Before proceeding to the current debate, a brief discussion is required beforehand regarding historical export controls placed over cryptography.

\subsection{Cryptographic Legacy}\label{section: cryptographic legacy}

The early history of cyber security export controls unfolds largely as a history of export controls placed by the U.S.~for cryptography. For a good part of the 20th century, the globally relevant export controls for cryptography were placed by the U.S.~with a dual rationale that balanced foreign policy with trade policy \citep{DiffieLandau07}.\footnote{~A number of historical examples could be used for illustrating this rationale. It was present when in 2008 when President Bush announced to reform dual-use export control policies \citep{Bartlett08}, to give a relatively recent example. To give another, more distant example: in the long historical picture, the rationale was visible already with the U.S.~reluctance for engagement in the early multilateral efforts, as exemplified by the late 1935 ratification of the (1925) Geneva Arms Traffic Convention \citep{Erickson15}. To some extent, both examples could be used as evidence for framing even the whole history of modern export controls against the dual rationale.} After the 1990s ``crypto wars'' \citep{Buchanan16b, Landau14, Winkel03}, these controls for  cryptography were largely and incrementally liberalized during the late 1990s and early 2000s within the transatlantic block. According to \citet{DiffieLandau07}, in the long haul, the policy change in the U.S.~was a result from a number of large transformations; the end of the bipolar world order, the rise of ICT and the associated demand for high-technology dual-use components, the simultaneous deregulation of the financial and ICT sectors among the Western states, the ever continuing globalization, and a number of related major historical trends. All in all, it was increasingly difficult to control the volatility of cyber security technologies, including but not limited to cryptography.

For a short while, it seemed that the old Cold War era regulations would prevail for cryptography. The WA  was amended in 1998 to cover ``strong'' encryption (as measured by a key length of a cipher), albeit shorter keys were framed to outside of the WA's scope by a request from the U.S. and its ICT sector \citep{Dean99, DiffieLandau07}. Paved by the scientific evidence in favor of the insecurity of many weak cryptographic algorithms -- among other things -- the efforts of the Clinton administration failed at both domestic and international venues \citep{Buchanan16a, Dean99, Thomsen01}. Soon after these failed attempts, the historical policy regime for cryptography was in practice terminated by the European Union in 2000 with its abolition of the cryptographic export controls within the \text{union -- a} decision that was later adopted also by the Clinton administration \textit{vis-\'a-vis} the fifteen EU members together with a number of other Western countries  \citep{DiffieLandau07, McGlone00, Thomsen01}. Although some regulations over exports of cryptography are still in place in the U.S., a more recent trend seems to include further relaxation and small amendments rather than a push toward tighter controls \citep{Bartlett08, Landau14}. For the controlled cryptography exports, different licensing choices are available to countries outside of the WA. Similar deregulation occurred also in other ICT subdomains.

At the same time, this liberalization has gone hand in hand with the expansion of multilaterally covered items. In other words, there has been a parallel trend for covering more and more technology items as the participating states have attempted to keep up with the pace of technological change. This dual trend of expansion and relaxation applies also to the multilateral case at hand: the WA's plenaries from the early 2000s to 2015 tend to emphasize both relaxation and expansion \citep[see][for instance]{Wassenaar00, Wassenaar02, Wassenaar15}. The latter is understandable against the noted rationale of achieving efficiency through comprehensiveness and diligence. Finally, it should be emphasized that the historical crypto wars also left behind an energetic tradition of active NGO involvement, driven by cyber security professionals, lawyers, academics, open source advocates, and hobbyists alike.\footnote{~Although some scholars have interpreted the EU response as having resulted from opaque U.S.~policies for cryptography \citep{Winkel03}, it is worthwhile to emphasize this NGO-side, including the role of academia. Indeed, perhaps the most famous legal case during the crypto wars was initiated by computer scientist Daniel J.~Bernstein -- a case that can be interpreted as having signaled the fundamental turn of tide \citep{Kennedy00, McGlone00, Thomsen01}. Analogously, emergence of the open source phenomenon accelerated the relaxation \citep{DiffieLandau07}. Although the paper is framed against multilateral controls (policies), which are imposed by states against other states, it should be thus emphasized that export control politics cannot be understood with a state-centric viewpoint alone.}

\section{The Wassenaar Debate}\label{section: wassenaar}

The discretion principle, the balance between security and commerce, and the crypto wars are all helpful for better understanding the 2010s Wassenaar debate. However, these are not enough; cyber security cannot be fully understood without understanding technology. Thus, in what follows, a brief technical overview is carried out before revisiting the more recent political developments.

\subsection{Software Restrictions}

Analogously to the Australia Group, which maintains a comprehensive list of toxics as well as human, animal, and plant pathogens \citep{Atlas08}, Wassenaar currently includes nine categories with meticulously collected but still vaguely defined items. On the side of computer hardware, which occupies the categories of computers and electronics, the list includes everything from microprocessor microcircuits and microcomputer microcircuits to analogue-to-digital converters, static random-access memories, field programmable logic devices, and circuits designed for signal processing, not forgetting ``custom integrated circuits for which either the function is unknown or the status of the equipment in which the integrated circuit will be used is unknown'' \citep[p.~50]{Wassenaar18a}. Although many of the items are listed with specific operational ranges, the definitions are still encompassing and imprecise. In general, this vagueness reflects the increasing problems with the concept of dual-use technologies \citep{EilstrupSangiovanni18, Rath14, Rajeswari98, WilliamsJones14}. Thus, to see the forest from the trees, it is better to again focus on the historical changes and amendments, augmented by the fundamental framings and exclusion criteria used in the Wassenaar Arrangement.

The current list starts with an important reservation about software that is not covered. This exclusion provision is threefold. To quote from the current list \citep[p.~3]{Wassenaar18a}, the WA does \textit{not} cover items that fall to the following categories:
\begin{enumerate}
\item{Software either (a) available to the public via commercial (1) ``over-the-count\-er'', (2) ``mail order'', (3) ``electronic'', or (4) ``telephony call'' transactions with ``retail selling points'', or (b) which is designed ``for installation by the user without further substantial support by the supplier''.}
\item{Software that is ```in the public domain'; \underline{or}''}
\item{Software implementations related to otherwise authorized exports in case these involve object (that is, binary) code needed for ``the installation, operation, maintenance (checking) or repair''.}
\end{enumerate}

The first provision essentially excludes the commercial software industry. The emphasis on the four different transfer channels reflects the 2001 agreement that transfers should be permitted for intangible technology items, including software, regardless of the means by which a transfer takes place \citep{Wassenaar01}. Moreover, the exclusion of open source is effectively made explicit in the second exclusion criterion through a familiar public domain loophole \citep{Beck00, Burke12}. The third clause is also important: this exclusion criterion essentially means that the controls do not apply to software updates, including security patches to known and publicly disclosed vulnerabilities. Though, the third category leaves it unclear (i)~whether the exclusion applies only to binary updates, which is not necessary the case with open source software and transactions related to vulnerability information. It is also unclear (ii)~whether software for vulnerability and exploitation scanning could be classified as being necessary for software maintenance, checking, or ``repair''. Likewise, (iii) it seems fair to say that most (but not necessarily all) vulnerabilities and exploits are disclosed and published as-is -- without any particular ``support by the supplier''. Consequently, also vulnerabilities and exploits could be interpreted as \text{excludable -- insofar} as these are understood as abstract software defects with security implications and concrete software implementations to exploit these defects, respectively, especially when these are published under the public domain principle. 

However, importantly, the first clause and the third clause both carry a remark that any software related to information security is not applicable \citep[p.~3]{Wassenaar18a}. Through this loophole, in principle, any software security artifact can be potentially covered in the WA via relaxing the definition of software with respect to information security.

\subsection{Surveillance Items}\label{subsec: surveillance items}

Information security is placed in the WA as a second subcategory of computers, to accompany the other subcategory of telecommunications. The latter covers anything from underwater equipment, acoustic carriers, explosion devices, and jamming equipment to conventional Internet Protocol (IP) networking. The last item is also relatively new. The WA's expansion toward IP-based networking occurred between 2012 and 2013, and this expansion was also rapidly adopted to the EU regulations~\citep{BauerBromley16, Bohnenberger17, Wassenaar12}. For the present purposes, particularly noteworthy are the covered IP-based technologies, which include ``surveillance systems or equipment'' that satisfy \textit{all} of the following conditions \citep[direct quotations from][p.~87]{Wassenaar18a}:

\begin{enumerate}
\item{Communication technology performing the following three functions on a ``carrier class IP network'': (a) ``analysis at the application layer''; (b) ``extraction of selected metadata and application content''; and (c) ``indexing of extracted data''.}
\item{Communication technology that is designed to perform the following two functions: (a)~data extraction according to ``hard selectors''; and (b) ``mapping of the relational network of an individual or of a group of people''.}
\end{enumerate}

To disseminate the list, the function (a) in the first clause refers to communication technology designed for the so-called deep packet inspection operating at the seventh layer in the Open Systems Interconnection (OSI) model. The function (b) clarifies that only systems that harvest specific applications (related to, say, the domain name system, the hypertext transfer protocol powering the current Web, or the protocols governing electronic mail) are covered. Finally, the indexing function (c) might be interpreted to restrict the scope to systems that utilize a database, although the intention remains generally unclear for this particular clause. 

Thus far, in theory, the preceding list covers numerous network security products related to intrusion detection and prevention, whether commercial or open source, as well as the academic research domain related to the use of machine learning for such products. However, there is an important restriction present already in the first clause: the term ``carrier class IP network'' is used to refer to ``national grade IP backbone'', which effectively rules out conventional (layer-7) detection and prevention systems designed to guard smaller networks, including local area networks. The second clause continues the same theme, further framing the covered systems to those involved in harvesting ``hard selectors'' that identify individuals or groups  via their real names, electronic mail addresses, phone numbers, group affiliations, or street addresses. \citep[p.~87.]{Wassenaar18a} This emphasis is understandable due to the increasing importance of meta-data for security intelligence and surveillance technologies \citep{Landau16a}. Moreover, the first clause contains a note that excludes systems specifically designed for network quality of service or marketing purposes \citep[p.~87]{Wassenaar18a}. Insofar as common cyber attacks such as denial of service (or tools thereto) are interpreted to always involve network quality of service characteristics, this note is somewhat ambiguous, however.

All in all, the expansion toward IP-based surveillance systems can be seen to target deep packet inspection technologies that are designed to operate at a nationwide level. Due to the provisions outlined, the scope is relatively narrow \citep{Bohnenberger17}. The clear intention is to cover those technologies used for so-called mass surveillance.

However, the subcategory of information security is largely still reserved for cryptography. In fact, cryptography is used to frame all items that fall to the categories of ``non-cryptographic information security'', ``defeating, weakening or bypassing information security'', ``test, inspection and production equipment'' related to security, and information security software \citep[pp.~97--98]{Wassenaar18a}. Although the wording about defeating, weakening, and bypassing is alarming in terms of research and common security tools, the origins of the current debate are located not in the telecommunications items and surveillance technologies, but in a seemingly simple but far-reaching adjustment made to the definition of software within the computer category in the \text{Wassenaar~Arrangement}.

\subsection{Intrusion Software}\label{subsec: intrusion software}

The origins in the current debate trace back to the 2013 expansion of covered items with software ``specially designed or modified to avoid detection by 'monitoring
tools', or to defeat 'protective countermeasures', of a computer or network-capable device'' \citep[p.~221]{Wassenaar18a}. In the current item list, intrusion software is clarified to contain software implementations performing \textit{any} of the  functions enumerated in the following two clauses \citep[as previously, direct quotations from][p.~221]{Wassenaar18a}:

\begin{enumerate}
\item{``The extraction of data or information, from a computer or network-capable device, or the modification of system or user data; \underline{or}''}
\item{``The modification of the standard execution path of a program or process in order to allow the execution of externally provided instructions''}.
\end{enumerate}

An enforcement of these two cases would apply to large portions of security industry and computer science. Both clauses are all-embracing, covering in practice anything from intrusion detection software to automatic vulnerability scanners, including \textit{fuzzing} and other types of runtime memory analysis. The wording in the first clause contains some hints that the attempt might be to emphasize the integrity aspects of cyber security, but the wording about system modifications is still too imprecise for making definite conclusions. The first clause may also conflict with the noted exclusion criterion for software updates. Further examples are not difficult to pinpoint; the first clause leaves it unclear whether the research and practice of computer and network forensics is potentially covered, for instance.

The same ambiguity applies to the second clause, which may cover various debugging tools commonly used in software development. Implicitly, the clause seems to refer to exploits for common software vulnerabilities such as buffer overflows. If the intention was to include all exploits, on the other hand, the definition is not comprehensive enough. That is to say, the exploitation of many vulnerabilities does not require the execution of externally provided \text{instructions -- actually}, the arrival rate of such vulnerabilities has probably decreased over the years, owing to mitigative solutions, the use of memory-safe programming languages, and the ever increasing amount of different web vulnerabilities. Finally, many common software vulnerabilities are plain logical defects, which makes it practically impossible to define the concept of externally provided instructions.\footnote{~A simple example would be a structured query language injection; what separates commands externally injected to execution from other commands externally inserted to execution?}

The list of covered security items does not stop here. In fact, protective countermeasures are specifically noted to include data execution prevention, address space layout randomization (ASLR), and sandboxing. Because technologies such as ASLR had been available in most operating systems by 2013, it is thus clear that the amendment targeted also more sophisticated exploitation techniques via which ASLR might be countered. The same applies to sandboxing, which is used in many current web browsers, for instance. Analogously, the WA specifically lists antivirus (AV), endpoint, and personal security products, as well as intrusion detection systems (IDS), intrusion prevention systems (IPS), and firewalls as monitoring tools. \citep[p.~222.]{Wassenaar18a} Paradoxically, many of these monitoring tools also fall to the domain of the first clause; a firewall extracts data from a network, an AV product from a computer, and so forth.

Interestingly, loopholes were still reserved for some industries and technologies. In particular, the definition of intrusion  software does \textit{not} apply to the following three groups of software \citep[p.~222]{Wassenaar18a}:
\begin{enumerate}
\item{Debuggers, virtualization hypervisors, or software reverse engineering tools;}
\item{Software implementations for digital rights management (DRM);}
\item{Software that is installed by manufacturers, administrators, or end-users for ``the purposes of asset tracking or recovery''.}
\end{enumerate}

It thus seems that virtualization technologies might be excluded via the specific hypervisor exception. Debugging and reverse engineering seem to be also exempted, despite of the earlier clauses over the modification of the execution of a software program and the data extraction from computers and networks. Of course, DRM software was also excluded, which, if nothing else, signals that at least not all of the commercial voices were ignored during drafting of the amendment.

\begin{figure}[th!b]
\centering
\includegraphics[width=\linewidth, height=12.0cm]{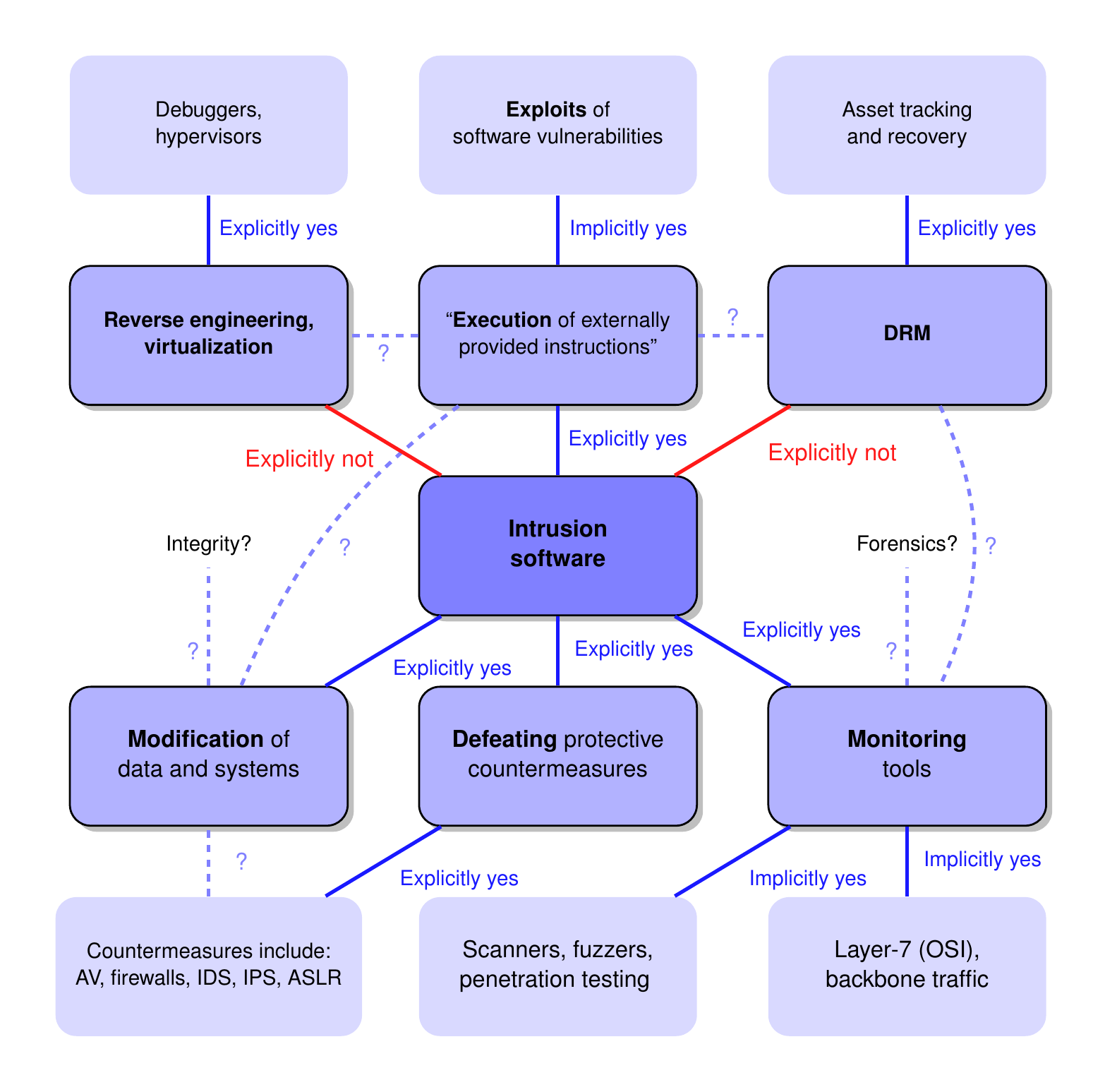}
\caption{\small The Fuzzy Analytical Meaning of Intrusion Software During the 2010s Wassenaar Debate (inferred from \citealt[][]{DepartmentOfCommerce15a} and \citealt{Wassenaar18a})}
\label{fig: intrusion software}
\end{figure}

For summarizing the key observations and ambiguities, an analytical conceptual model is presented in Fig.~\ref{fig: intrusion software} for the key terms related to intrusion software. In essence, the blurry definition of intrusion software underlines modification of data and systems, defeating protective countermeasures commonly used in defensive security, and monitoring of computers and networks. Implicitly, these characteristics relate to regulation of software vulnerabilities and exploits -- as well as surveillance technologies. The apparent inconsistencies in the figure allow to also understand how minor terminological changes can have major implications in export control regimes \citep[cf.][]{Rath14}. A minor terminological alteration prompted also the current, still ongoing Wassenaar debate. 

\subsection{Later Developments}

The 2013 amendment to the WA did not alter the fundamental discretion principle. Even though the multilateral amendment itself generated only limited media attention, the uproar started in May 2015 when the U.S.~Department of Commerce announced its proposal for implementing the elaborated surveillance and intrusion items. The proposal essentially altered the definition for intrusion software with a seemingly simple clarification. ``Systems, equipment, components and software specially designed for the generation, operation or delivery of, or communication with'' was interpreted to include ``network penetration testing products that use intrusion software to identify vulnerabilities of computers and network-capable devices'' \citep{DepartmentOfCommerce15a}. This clarification pushed the industry context explicitly toward the vulnerability markets and the offensive security industry segment.

Soon after, \citet{Google15}, a spokesperson from Google, commented that the rules were dangerously broad and vague, potentially causing an extensive financial licensing burden for companies that orchestrate vulnerability finding campaigns for improving the security of their products. On the NGO side, the Electronic Frontier Foundation (EFF), which had activated already during the mid-1990s crypto wars \citep{Kennedy00}, was also quick to respond, submitting also comments for the Department's request of comments. In addition to the economic arguments, EFF expectedly emphasized the relation to computer science and security research. In particular, EFF lifted \citep{EFF15} the additional definition that the ``technology for the development of intrusion software includes proprietary research on the vulnerabilities and exploitation of computers and network-capable devices'' \citep{DepartmentOfCommerce15a}. Given the keyword of \textit{proprietary research}, it is hardly a surprise that fierce criticism was expressed particularly in the offensive security industry segment.

\enlargethispage{1cm}

By July 2015, Synack, FireEye, and other companies in the segment had formed a coalition to oppose the proposed implementation, criticizing the ambiguity, the lack of attention given for the interests of U.S.~companies, and the negative impacts on global research and development. Regardless whether the change was a result from the opposition, or whether the proposal was intentionally published by the Department of Commerce as a hurried request of comments, the proposal was withdrawn later during the same month. The intense lobbying that followed made the U.S.~to reverse its position on the earlier WA amendments.\footnote{~This change allows to also make a small historical parallel to the ATT negotiations during which the U.S.~also eventually reversed its position \citep[see][]{Bromley12, Erickson15}. Another reversal occurred in April 2019 when the Trump administration announced a withdrawal from the ATT to which the U.S.~had joined (without ratification) in 2013 during the Obama administration.} In fact, the U.S.~representatives renegotiated the 2013 amendments at the December 2016 Wassenaar plenary session. To some extent, the renegotiation was successful: the current item list explicitly excludes vulnerability disclosure and (cyber) incident response \citep[p.~80]{Wassenaar18a}. While vulnerability disclosure, incident response, and associated factors were also the primary concerns of many industry stakeholders, the current exemption is again largely a line in the sand: depending on the interpretation and circumstances, vulnerabilities and exploits may be exempted or these may satisfy the definition given for intrusion software. Although the U.S.~sought to eliminate the concept of intrusion software altogether~\citep{Lichtenbaum18}, the concept still appears in the current item list. At the time of writing, it also seems likely that the concept will remain firmly as an item covered in the Wassenaar Arrangement.

\section{Discussion}

This paper reviewed the 2010s debate on augmenting the Wassenaar Arrangement to cover cyber weapons. The elaborated discretion principle and the history of cryptography export controls together with the brief technical overview help at digesting the numerous different points raised during the debate. In addition to summarizing the key observations, the discussion that follows enumerates a few tentative policy recommendations.

In some respects, the current debate resonates with the 1990s debates on export controls for cryptography. During that time, there was an increased awareness among policy makers for regulating cryptography via export controls, which was accompanied by intense lobbying and widespread confusion on how the actual regulation should work. In both cases, state institutions seem to have failed in attracting professionals and volunteers with sufficient domain knowledge for participating in policy drafting \citep[cf.][]{Beck00, Dean99}. In this area concrete improvements might be also relatively easy to achieve. In other words, it is easy to agree with a commonly voiced \citep[e.g.,][]{BauerBromley16, HumanRightsWatch14, Shaw16} policy recommendation (PR) that can be stated \text{as follows}:

\begin{enumerate}[label=PR$_\arabic{enumi}$]
\item{Encourage active participation of diverse stakeholders in the policy making of cyber security export controls.}\label{pr: stakeholders}
\end{enumerate}

When compared to export controls for conventional arms and weapons, the current debate and the historical crypto wars are also relatively unique in the sense that NGOs such as EFF and the supply-side industry have mostly joined forces to oppose state initiatives particularly in the United States. When compared to the crypto wars, the historical context is different, however. In addition to the substantially increased importance of cyber security, the present attempts occurred after a relatively long period of relaxed controls, owing to the deregulation that was implemented in the aftermaths of the crypto wars, which, in turn, were fought under a historical legacy of tight export controls. There are also additional reasons why the historical questions related to cryptography are relevant for framing the still ongoing Wassenaar debate.

It is interesting to observe that human rights activists have almost universally promoted the unrestricted use of cryptography and the free exchange of strong encryption technologies. This enduring consensus largely originates from the crypto wars. In contrast, there has been some polarization among the viewpoints of NGOs during the Wassenaar debate. While the position of EFF has aligned with those held in the cyber security industry, other NGOs have promoted the establishment of export controls for offensive cyber weapons. In particular, human rights groups had a strong influence over the 2013 inclusion of intrusion software to the Wassenaar Arrangement \citep{BauerBromley16, Bohnenberger17, Burkart17}. This polarization reflects a wider crack. Because the concept of intrusion software was included verbatim in the 2018 amendment to the EU regulations~\citep{EC18a}, while the analogous implementation attempts seem to have stalled in the U.S.~\citep{Lichtenbaum18}, it seems that the approaches to cyber security regulation continue to diverge among Western states also in terms of export controls.\footnote{~It can be remarked that the mid-2010s events in the EU largely followed a similar path, although the analogous European lobbying efforts were not as successful as in the United States. In the late-2010s, however, there were also some cracks within the EU. Although the attempts were unsuccessful, some member states (notably, Finland, Sweden, and the United Kingdom) opposed the EU-level enforcement by arguing that regulation should occur primarily through the WA and its national implementations \citep{Mobbrucker18}.} Analogous trends are seen in other cyber security domains, including but not limited to privacy and data protection. It remains interesting to see whether and to which extent these diverging trends will impact multilateralism in general.

The historical controls for exports of cryptography balanced security concerns with economic interests. Although the voice of the cyber security industry eventually reached the right target, the 2015 proposal of the U.S.~Department of Commerce still marked an abrupt albeit temporary change in the continuum between security and trade. Because exporting companies need to devote substantial amount of time and resources to study changes in export control policies, preemptive policy changes are generally preferable \citep{Seyoum17}. Changes to export control policies should not come out from the blue sky to industry stakeholders. Thus, it is important underline the adjective \textit{diverse} used in \ref{pr: stakeholders}. In other words, also industry stakeholders should be encouraged to participate in the drafting of cyber security export control policies.

It is important to emphasize that despite of some NGO-driven surveys and academic studies  \citep{Burkart17, McKuneDeibert17, Ruohonen16RCIS, Wilson13}, the existing knowledge about the offensive cyber security segment remains extremely limited for stakeholder-accommodating but still evidence-based policy making. It is also unlikely that the situation will improve without industry participation. Therefore, also other means should be considered to motivate voluntary industry participation. As alternatives to regulation, incentives could be created for promoting good corporate governance practices, ethical codes of conduct \citep{Cavelty14, WilliamsJones14}, and the establishment of industry consortia in order to improve transparency and accountability. These are examples that could accommodate also some of the human rights concerns. Without neither referring to any particular policy technique nor answering to critics \citep{Parsons17}, these softer (non-regulatory) aspects are implicitly embedded to the policy recommendation \ref{pr: industry}:
\begin{enumerate}[label=PR$_\arabic{enumi}$, resume]
\item{Improve outreach to the cyber security industry.}\label{pr: industry}
\end{enumerate}

The heritage from the crypto wars can be portrayed also through an institutional viewpoint to export controls. In this regard, the Wassenaar debate reflects the long-standing institutional issues between different U.S.~state departments and bureaus for handling export controls.  Comparable problems are seen also in Europe. In essence, the continuing departmental infighting and related institutional conflicts in the United States \citep{Seyoum17, Waltz07} manifest themselves within the EU through cross-country harmonization and enforcement problems \citep{Micara12, vonderDunk09}. Reflecting the commonly voiced urge to have central clearinghouses \citep{Burke12, Burkart17, Seyoum17}, the following policy recommendation seems generally justified:
\begin{enumerate}[label=PR$_\arabic{enumi}$, resume]
\item{Improve centralized governance of cyber security export controls.}\label{pr: governance}
\end{enumerate}

Cryptography cannot be fully understood without deep scientific knowledge. In contrast to mathematics and algorithms, the discussion about cyber weapons continues to revolve around vague abstractions that cause numerous ambiguities. In this respect, the language used in the WA and related policy documents foretell about limited participation of relevant stakeholders. Indeed, like with many related cyber security questions \citep[][]{BellovinLandau16}, the vetting of the amendments was limited among cyber security professionals and computer scientists. Despite of some improvements, particularly the engagement with academia remains generally limited \citep{BauerBromley16, Evans12}, which further reinforces the importance of diverse stakeholder participation (\ref{pr: stakeholders}). Wider participation implies slower policy making, which may be also beneficial in the cyber security domain. The sequence of events in the Wassenaar debate indicate that the amendments and proposals were likely made too hastily.\footnote{~As noted by \citet{EilstrupSangiovanni18}, the speed at which multilateral agreements and conventions are negotiated has been a common issue also in the cyber security domain. The same point extends to the right timing to conduct such negotiations in the face of other events in world politics. In other words, it may be that mid-2010s was not the optimal time to alter the WA.} Because changes once made to multilateral export control systems are difficult to revert or otherwise alter, care should be taken with further amendments, as summarized in the policy recommendation~\ref{pr: diligence}:

\begin{enumerate}[label=PR$_\arabic{enumi}$, resume]
\item{Balance comprehensiveness with diligence in each and every amendment.}\label{pr: diligence}
\end{enumerate}

There are three additional points worth making about terminology. As was demonstrated in Section~\ref{section: wassenaar}, even a cursory technical outlook still reveals many unclear and contradictory issues. These problems prompt the first practical point: because offensive security tools and techniques are imposed by regulations placed over payment card industry \citep{Knowles16}, for instance, the ambiguity issues may lead to legal conflicts and other problems for a national enforcement of the WA. This is the main message from the 2015 fiasco in the United States.

The second point is more theoretical: due to the vagueness, it is difficult to infer about the actual intention of policy makers from the export control policy documents. To make the point more  translucent, it is illuminating to briefly return to Fig.~\ref{fig: intrusion software}. From a technical perspective, there are numerous questions related to the edges (relations) between the vertices (concepts) shown in the figure. In many respects, it would make sense to redraw the figure as a maximally connected graph by connecting all of the rectangles together. Is fuzzing reverse engineering? Is asset tracking part of monitoring? How about surveillance? Such questions are important to ask because imprecise terminology poses a risk of escalation; legitimate attempts to control exports of state-led surveillance technologies may spill over to industry and civil society.\footnote{~It is not difficult to ask even trickier questions. By recalling the definitions from Section~\ref{section: wassenaar}, consider the following imaginary question as an example: is the use of the \textit{nmap} tool \citep[cf.][]{Nmap09} part of \textit{monitoring} or (and) \textit{defeating of countermeasures}, such as \textit{firewalls}, when conducted in, say, \textit{virtualized} environments for \textit{penetration testing} of an \textit{open source} software \textit{updating} solution deployed for a particular \textit{hypervisor} running in a cloud?} Fortunately, it is not necessarily difficult to prevent such an escalation. While keeping in mind that precise language often falls on deaf ears due to bargaining and other political reasons affecting export controls \citep{Gartner10, Hansen16}, addressing the following policy recommendation (\ref{pr: terminology}) would offer a good start for concrete improvements:
\begin{enumerate}[label=PR$_\arabic{enumi}$, resume]
\item{Rely on established terminology used in the cyber security industry and computer science.}\label{pr: terminology}
\end{enumerate}

This recommendation can be accompanied with a corollary.\footnote{~Also \citet{Herr16} raise the same point about \ref{pr: terminology}. It is worth further remarking that a robust way to improve the terminology would be to attach the concepts to standards, but the problem is that particularly offensive cyber security tools and techniques seldom have clear technical reference points.} The Wassenaar debate supports the arguments that the concept of dual-use has already long ago lost most of its analytical power \citep{Burkart17, EilstrupSangiovanni18, Rath14, Rajeswari98, WilliamsJones14}. If it is admitted that cyber security differs from conventional security, it is also logical to make a subsequent policy recommendation:

\enlargethispage{1cm}

\begin{enumerate}[label=PR$_\arabic{enumi}$, resume]
\item{Review the concept of dual-use in the context of cyber security.}\label{pr: dual-use}
\end{enumerate}

Without addressing the recommendation \ref{pr: dual-use}, there is a continuous danger that more and more cyber security technologies will be moved to the other endpoint in the civilian-military continuum. The consequences are unforeseeable, but there are still many good reasons to assert that such changes would have a deteriorating effect upon security in the Internet.

The third and final point about terminology is speculative. To some extent, it seems that the language used in the export control policy documents is intentionally used for trying to conceal the elephant in the room; the connection of export controls to government hacking and state-led offensive security in general. By using the obscure term intrusion software, states can conveniently avoid mentioning many inconvenient cyber security questions. These questions include legislations about planting malware to citizens' computers \citep{BellovinLandau16}, requiring private sector ICT companies to hand over sensitive data, the potential of inserting ``golden keys'' to cryptographic algorithms \citep{Buchanan16a}, the intentional introduction of vulnerabilities, and, of course, the suspected heavy use of the offensive cyber security industry segment for purchasing vulnerabilities and exploits \citep{Burkart17, Deibert15, Jardine15, KuehnMueller14a}. If anything, the fundamental lesson learned from the crypto wars was that restricting technology and holding secrets make the whole Internet a less secure place, irrespective whether the restrictions or secrets refer to weaknesses in cryptographic algorithms or to undisclosed software vulnerabilities.

The fundamental problem for academic research is that most allegations of government hacking remain speculations that are better done by journalists, cyber security companies, NGOs, and related parties. The point applies also to the Wassenaar debate. Countries such as Saudi Arabia, Libya, and Syria are not participating in the Wassenaar Arrangement (see Fig.~\ref{fig: wassenaar}), but it is extremely difficult to establish a robust link between government hacking and related exports to such countries. For these reasons, there is no other option but to resort to the final policy recommendation~\ref{pr: transparency}:

\begin{enumerate}[label=PR$_\arabic{enumi}$, resume]
\item{Improve government transparency in the cyber security domain.}\label{pr: transparency}
\end{enumerate}

The reverse direction should be also considered; there are no exports without imports. Because countries such as China and Israel remain outside of the international regulation regime, it remains unclear how effective the export controls for cyber weapons are due to imports from outside of the regime. There is also a risk that non-state actors may be able to import surveillance technologies and knowledge thereto. After all: there are already plenty of alarming examples about leakages of cyber weapons from various state actors. 

The points raised can be used to briefly reflect the bigger issues working underneath the ongoing debate about export controls for cyber weapons. Cyber security is often framed with different polar opposites, such as external and internal, military and civilian, or public and private~\citep{Ruohonen16GIQ}. The Wassenaar debate acts as a good reminder that there are also many other fundamental dimensions. As was elaborated, exports controls in general align well with the nexus between trade and security. This nexus is not the only one, however. Framing can be done also in terms of conflicting interests between states and industry, between states and civil societies, between national and societal security~\citep{Mueller17}, or between industry and human rights. It is possible to go even further. As \cite{Schneier19} has been arguing, one fundamental question is about whether the future is built around security or surveillance; either ``everyone gets to spy, or no one gets to spy''. When portrayed through the Wassenaar Arrangement, it seems fair to maintain the sensibility of his point: it indeed seems improbable that the arrangement can prevent the proliferation of surveillance technologies and intrusion software, whatever the exact definitions for these are. This tentative prediction echoes the historical voices from the crypto wars.

When the nexus is set between security and surveillance, the Internet and its governance are necessarily also a part of the picture. In this regard, the final analytical dimension can be set according to \citeauthor{Mueller10}'s \citeyearpar{Mueller10} classical framing between networked multistakeholder and state-centric governance models. Multilateral export controls set by states against other states are a \textit{prima facie} example of the latter. Although vulnerability disclosure and incident response are currently exempted from the WA's coverage, the 2010s debate and the arrangement's enforcement particularly in the EU indicate another small swing in the pendulum toward state-centric governance. A drop in the ocean, perhaps, but from small streams becomes a river.

\enlargethispage{1cm}
\begingroup
\parindent 0pt
\parskip 4pt
\def\enotesize{\scriptsize}
\theendnotes

\bibliographystyle{apalike}

\end{document}